\documentclass[a4paper,11pt]{article}
\usepackage{pos}

\usepackage{cleveref}

\usepackage{braket}
\usepackage{bm}

\hypersetup{pdfinfo={
Title={Hadron spectroscopy and few-body dynamics from lattice QCD},
Author={Andrew D. Hanlon}
}}

\crefname{figure}{fig.}{figs.}
\Crefname{figure}{Fig.}{Figs.}
\crefname{table}{tab.}{tabs.}
\Crefname{table}{Tab.}{Tabs.}
\crefname{equation}{eq.}{eqns.}
\Crefname{equation}{Eq.}{Eqns.}
\crefname{listing}{lst.}{lsts.}
\Crefname{listing}{Lst.}{Lsts.}
\crefname{section}{sec.}{secs.}
\Crefname{section}{Sec.}{Secs.}
\crefname{appendix}{app.}{apps.}
\Crefname{appendix}{App.}{Apps.}

\title{Hadron spectroscopy and few-body dynamics from lattice QCD}

\author{Andrew D. Hanlon}

\affiliation{Physics Department, Brookhaven National Laboratory, Bldg. 510A, Upton, New York 11973, USA}
\affiliation{Department of Physics, Carnegie Mellon University, 
Pittsburgh, Pennsylvania 15213, USA}

\emailAdd{adhanlon@cmu.edu}

\abstract{Despite quantum chromodynamics (QCD) being established as the theory of the strong interaction and its many successes since then, significant challenges in our understanding of hadron physics remain.
The lack of a full understanding for how the observed hadrons arise from the quark and gluon degrees of freedom which define QCD represents a real challenge in connecting the theory to experiment.
In particular, the rich spectrum of hadrons marks a significant gap in our understanding, as no model can currently explain all observed hadrons and QCD itself has only been used to study a small subset of the full spectrum.
The significant hurdle for using QCD directly is due to the non-perturbative nature of the theory at low energies, requiring methods like lattice QCD.
Here, the spectrum of hadrons and how it relates to few-body dynamics is reviewed, with a focus on state-of-the-art methods for their study through lattice QCD.}

\FullConference{The 40th International Symposium on Lattice Field Theory (Lattice 2023)\\
July 31st - August 4th, 2023\\
Fermi National Accelerator Laboratory\\}

\begin{document}
\maketitle

\section{Introduction}

It has long been accepted that quantum chromodynamics (QCD) is the correct theory of the strong interaction.
In the 50 years since its formulation, QCD has been hugely successful at high energies where perturbation theory can be applied.
However, this rich and elusive theory still has much left to uncover.
A significant hurdle towards progress in strong-interaction physics is the low-energy regime of QCD where the theory becomes non-perturbative.
Additionally, the observed degrees of freedom are not the quarks and gluons which define QCD but rather composite states corresponding to color singlets.
How the underlying theory of QCD leads to the vast number of observed hadrons is still not understood and remains an active area of research.

Adding to the confusion, there have been questions raised regarding the reliability of certain states observed in the hadronic spectrum.
The overwhelming majority of hadrons are resonances (i.e. unstable states) which in many instances correspond to the observance of a bump-like behavior in the cross section.
However, there are other structures within the scattering amplitudes, not corresponding to resonances, that can cause similar behavior that may be misconstrued as actual resonances or may simply lead to incorrect resonance properties.
Fortunately, there are rigorous methods to reliably extract resonance information from poles in the scattering amplitudes.
These methods enforce certain constraints on the structure of analytically continued scattering amplitudes.

In addition to the use of these methods when analyzing experimental data, they can also be used in conjunction with theoretical calculations.
However, due to the difficulty in direct calculations of QCD at low energies, much of our theoretical understanding comes from various models.
The quark model in particular was a crucial step towards organizing and understanding the origin of the huge number of observed hadrons.
But, with the observation of many exotic hadron candidates in the past two decades that do not quite fit the quark model, several extensions have been needed to describe these new states. 

While models are important to build intuition, direct calculations within QCD would be extremely beneficial.
Lattice QCD can be used to reliably extract scattering information through relationships between the scattering amplitudes in infinite volume and the finite-volume spectrum.
Then, utilizing the more sophisticated analysis methods based on the correct analytic structures for the scattering amplitude, these results along with experimental methods constitute a powerful duo for the robust understanding of the QCD spectrum.

\section{The Spectrum of Hadrons}

The 1950s and early 1960s were an exciting, albeit confusing, time for particle physics with the discovery of many particles that did not fit expectations at the time.
With the advent of the hugely successful quark model~\cite{Gell-Mann:1964ewy,Zweig:1964ruk,Zweig:1964jf}, much of this confusion subsided for quite some time, as we then had a consistent means for categorizing the particles being seen in experiment and predicting ones yet to be seen.
However, this period was not to last, and we are again in the midst of a similar situation where many new particles are being discovered that do not fit any given model.

Thus, we once again find ourselves in need of improving our theoretical methods to explain the scattering data within the past few decades.
There has indeed been significant progress towards this from various directions (see e.g. Refs.~\cite{Mai:2021lwb,Mai:2022eur} for recent reviews on theoretical methods for hadron spectroscopy).
We focus here on the theoretical work relevant for calculations in lattice QCD.

\subsection{The Quark Model}

Up until the early 2000s, the quark model was able to reasonably describe all observed hadrons as baryons (three quarks) or mesons (quark and antiquark), which on the one hand was an immense achievement, but on the other hand it was a great puzzle as to why it worked so well.
In fact, QCD allows for many more types of states, so long as they are color singlets.
These extra states are referred to as exotics: e.g. glueballs (made entirely of gluons),  hybrid mesons (mesons with a gluonic degree of freedom), tetraquarks (two quarks and two antiquarks), pentaquarks (four quarks and an antiquark), etc.

\begin{figure}
    \centering
    \includegraphics[width=.25\columnwidth]{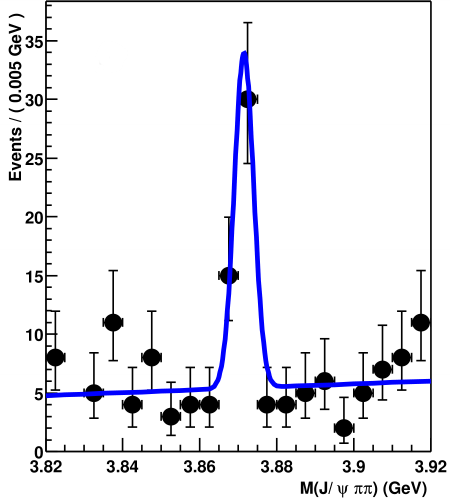}
    \includegraphics[width=.38\columnwidth]{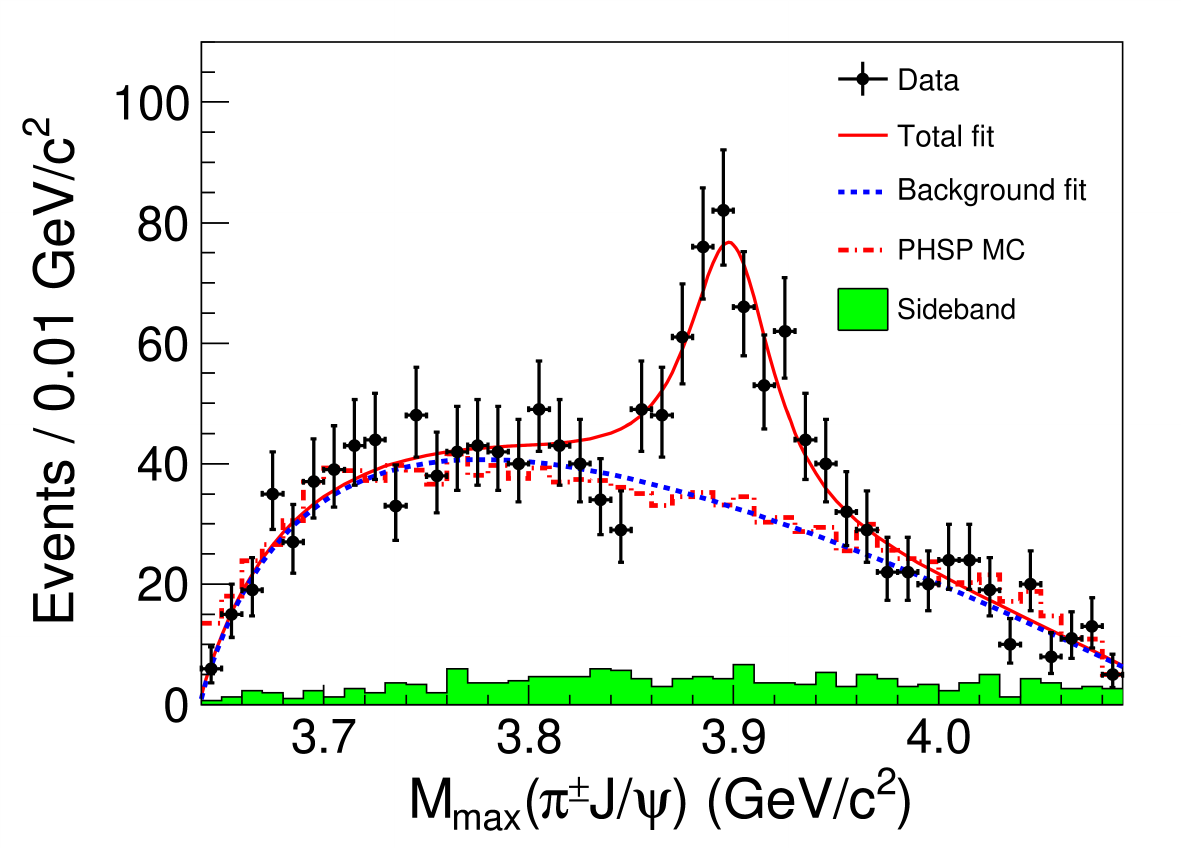}
    \includegraphics[width=.35\columnwidth]{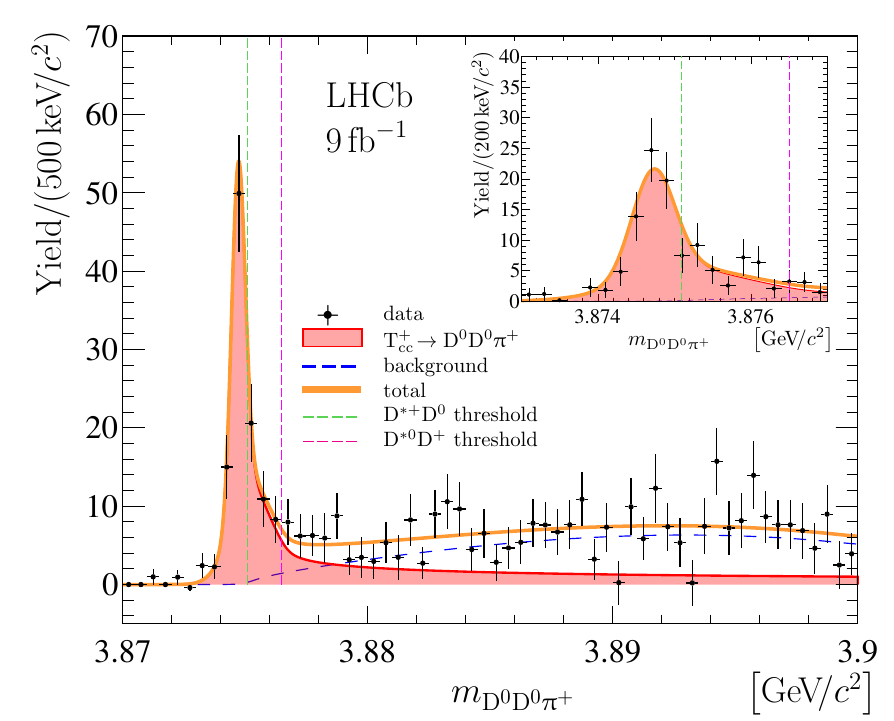}
    \caption{Selected examples of exotic candidates discovered in the past two decades. The $X(3872)$ discovered by Belle~\cite{Belle:2003nnu} which defined expectations from the quark model, the $Z_c (3900)$ discovered by BESIII~\cite{BESIII:2013ris} which is a charged charmonium state and thus must be more than $c \bar{c}$, and the $T_{cc}^+ (3875)$ discovered by LHCb~\cite{LHCb:2021vvq,LHCb:2021auc} which has two charm quarks and integer spin and therefore must be exotic.}
    \label{fig:exotics}
\end{figure}

However, starting with the discovery of the $D_s(2317)$ by BaBar in the spring of 2003~\cite{BaBar:2003oey}, a flood of new states, some of which were exotic candidates that did not fit the quark model, began to appear.
Three particularly interesting states, with their discoveries spread across 20 years, are shown in \Cref{fig:exotics}.
There are various reasons for some of these states not fitting the quark model.
For instance, some have quantum numbers that cannot be realized using only the conventional hadrons, some simply have properties that are very different than predicted from the quark model, and others may not correspond to any state predicted at all.

It should also be noted that prior to this preponderance of new states that could not be explained by the quark model, there were several cases in which the quark model was showing its limits.
For example, the quark model could not explain why the Roper resonance lies
below the lightest negative-parity nucleon, or why the $\Lambda(1405)$ is lighter than its nucleon counterpart.
Further, it was pointed out by Jaffe that a tetraquark description of the light scalar mesons fit better with experimental observations~\cite{Jaffe:2004ph}.

Despite its shortcomings, the quark model still plays an important role in our understanding of the hadronic spectrum by giving context to and intuition for the experimental observations.
Further, lattice QCD calculations typically use the quark model as a basis for the interpolating operators used in simulations.

\subsection{The Scattering Amplitude}

To address the issues with the hadronic spectrum, a good theoretical understanding of the scattering amplitude $T$ is needed.
The various states in QCD are understood to be each associated with a pole in the scattering amplitude.
The extraction of the pole and residue corresponding to a particular state is the most rigorous way to determine the properties of that state, as, unlike other methods for studying QCD states, the properties of these poles are process independent.
The scattering amplitude $T$ is defined in terms of the scattering matrix $S = 1 + i T$ which relates the \textit{in} and \textit{out} states\footnote{Several different conventions for normalizations exist in the literature. Here we follow those laid out in Ref.~\cite{Martin:1970hmp}.}
\begin{equation}
\begin{split}
    _{\rm out}{\braket{f|i}}_{\rm in} &\equiv \bra{f} S \ket{i} \\
                                      &= \delta_{fi} + \bra{\bm{p}^f \alpha^f} i T \ket{\bm{p}^i \alpha^i} \\
                                      &= \delta_{fi} + i \delta^4(p^f - p^i) \mathcal{T}_{\alpha^f \alpha^i}(\bm{p}^f, \bm{p}^i),
\end{split}
\end{equation}
where $p^i \equiv (E_i, \bm{p}^i)$ is shorthand notation for the set of four-momentum of the initial state, $\alpha_i$ specifies the scattering channel and the internal degrees of freedom (e.g. spin, isospin, etc.) of the initial particles, and similarly for the final scattering state.
Note that, although the scattering amplitude depends on the three-momentum of the scattering particles, symmetry considerations restrict the values they can take.
In particular, Poincar\'e invariance reduces the number of independent kinematic variables that the amplitude depends on by ten, corresponding to the ten generators of the Poincar\'e symmetry group.

In the case of two-to-two scattering, this leaves just two independent kinematic variables on which the scattering amplitude depends corresponding to the total energy $E_{\rm cm}$ and the scattering angle $\theta$ in the final state, which are both taken to be defined in the center-of-momentum frame.
Note that typically the coordinate system used in the center-of-momentum frame is taken to be such that the relative motion of the initial particles is aligned along the $z$-axis, and then the polar angles for the final state $\theta$ and $\phi$ take on their conventional definitions. 
Thus the symmetry in rotations about the $z$-axis leads to $\phi$ dependence in the scattering amplitude only as an overall phase.
Further, two-to-two scattering admits three scattering processes or ``channels'' with their individual center-of-momentum energy squared corresponding to the three Mandelstam variables\footnote{To go beyond two-to-two scattering, one can define generalized Mandelstam variables but this just makes the problem more complicated.}
\begin{align}
    s &\equiv (p^i_1 + p^i_2)^2 = (p^f_1 + p^f_2)^2 = E_{\rm cm}^2 , \\
    t &\equiv (p^i_1 - p^f_1)^2 = (p^f_2 - p^i_2)^2 = -\frac{E_{\rm cm}^2}{2} (1 + \cos \theta) , \\
    u &\equiv (p^i_1 - p^f_2)^2 = (p^f_1 - p^i_2)^2 = -\frac{E_{\rm cm}^2}{2} (1 - \cos \theta) .
    \label{eq:mandelstam}
\end{align}
These Lorentz invariant kinematic quantities are a convenient set to use for the dependence of the scattering amplitude, rather than the three-momentum.
Thus, in this case of two-to-two scattering, the scattering amplitude can be written as $\mathcal{T}_{\alpha^f \alpha^i}(s, t, u)$, where sometimes $u$ is excluded as it can be written in terms of $s$ and $t$.
However, some physical intuition can be gained through considering the scattering amplitude to depend on all three Mandelstam variables,
where we can consider the scattering amplitude to be living in a two-dimensional space spanned by three linearly-dependent coordinates corresponding to the Mandelstam variables, which shows their interdependence.
In particular, the possible physical scattering region for each channel places limits on the values the Mandelstam variables can take and these represent disjoint regions within the two-dimensional space.
In this way, as the physical scattering regions are disjoint, one can define a single scattering amplitude living in the entire two-dimensional space with the property that it becomes equivalent to the scattering amplitude for a particular channel when the Mandelstam variables take on the values corresponding to the physical scattering region for that channel.
Then, upon analytic continuation to complex values for the Mandelstam variables, one can provide relationships between the scattering amplitude in one channel and in the other two.
That the scattering amplitude indeed can be defined in this way is know as the Mandelstam hypothesis.
In all, there are three general properties that we demand the scattering amplitude to satisfy:
\begin{enumerate}
    \item Unitarity of the $S$-matrix - The squared matrix elements $|S_{fi}|^2$ are interpreted as the probability of an initial state $\ket{i}$ to evolve into the final state $\ket{f}$.
    From this interpretation, conservation of probability demands $\sum_f |S_{fi}|^2$ to be unity, which in turn implies $S^\dagger S = 1$, and therefore $S$ must be unitary.
    A consequence of unitarity is the occurrence of a branch cut starting from each threshold and going off to the right towards infinity.
    The first sheet, whose real axis is where the scattering amplitude can be measured, is referred to as the physical sheet.
    Thus, each new threshold doubles the number of Riemann sheets in which the scattering amplitude lives.
    \item Crossing symmetry - This is the relationship between the various scattering channels and is essentially the statement of the Mandelstam hypothesis discussed above.
    A consequence of crossing symmetry is the occurrence of a left-hand cut coming from the partial-wave projection of the scattering amplitude containing the pole associated with the exchanged particle or particles in the other channels.
    \item Causality - This implies the commutator of two fields must be zero if they are space-like separated. The consequences of this can be shown to require analyticity in certain regions of the scattering amplitude. This is generally used to demand the scattering amplitude only take on singularities that are associated with some physical origin.
\end{enumerate}
The analyticity constraints on the scattering amplitude limit the location for which poles associated with various states arising in the theory can be located.
Specifically, there are three types of poles allowed, each corresponding to a different type of state: i) poles on the real axis of the physical sheet below threshold correspond to bound states, ii) poles on the real axis of an unphysical sheet below threshold correspond to virtual bound states, and iii) poles off the real axis on an unphysical sheet correspond to resonances.
No other poles are allowed without violating the properties above.
It should be noted that there is really no clear distinction in the origin of the different pole types, despite the different names for each.
And, in fact, poles have been seen to evolve from one type to another as e.g. the quark masses are changed.

\begin{figure}
    \centering
    \includegraphics[width=.5\columnwidth]{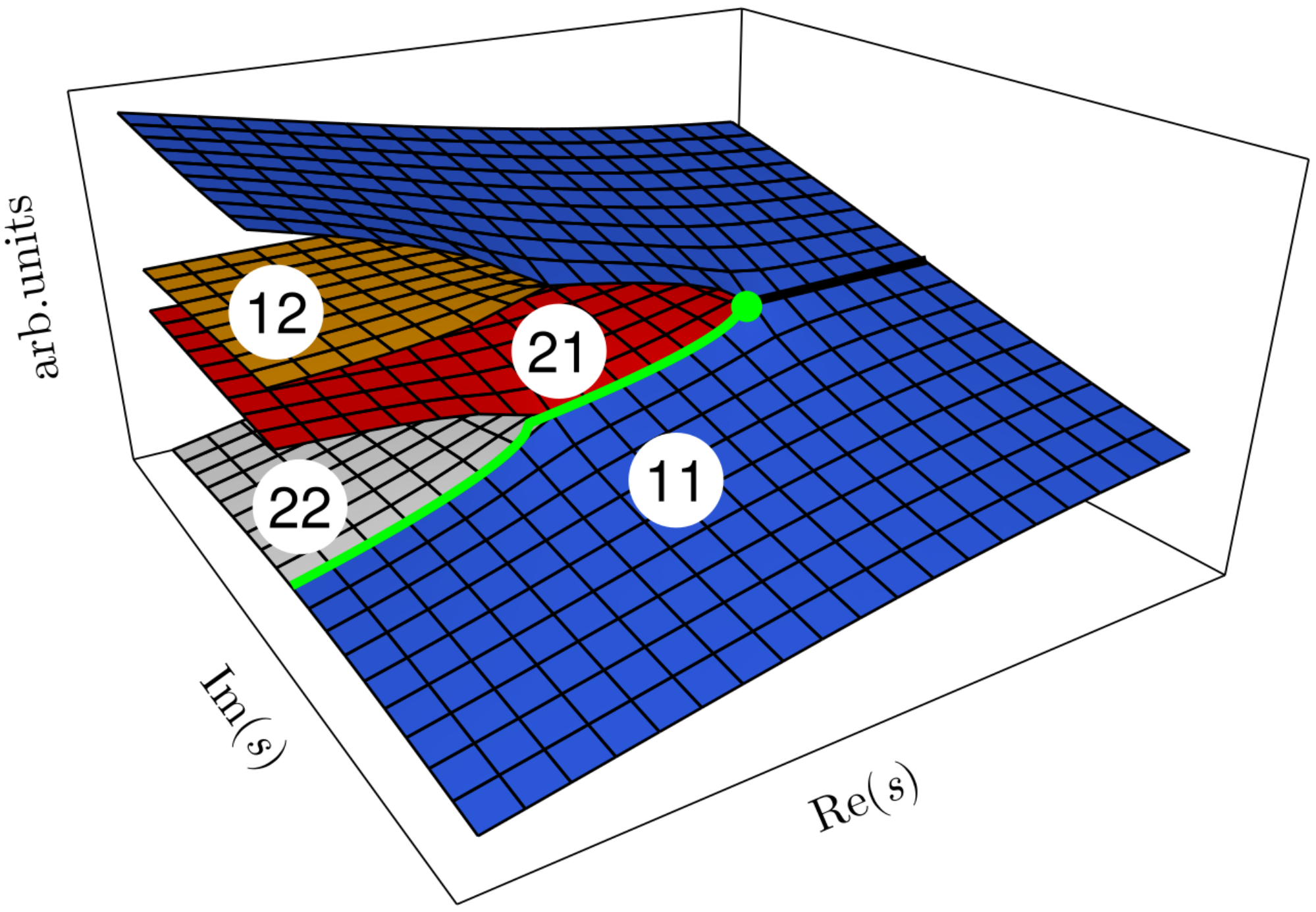}
    \caption{The four Riemann sheets corresponding to two open scattering channels.
    The green line corresponds to the location of physical scattering.
    Figure taken from Ref.~\cite{Workman:2022ynf}.}
    \label{fig:riemann_sheets}
\end{figure}

In the case of two open channels, the Riemann sheets produced are shown in \Cref{fig:riemann_sheets}.
This figure shows the difficulty faced when attempting to uncover poles that may be far from the region in which physical scattering takes place (represented by the green line).
For poles nearby, and with a clear path to, the region of physical scattering, the typical bump structure associated with a resonance is likely to be seen.
But what is not so clear, is how poles far away from this region will be seen (or not seen).
Additionally, for poles nearby the singularities produced in the crossed scattering channels, the effects of those singularities may mask the poles.
This makes including the correct analytic structure an essential aspect of a robust analysis in those cases.
However, the three general properties of the $S$-matrix described above are not always taken into account in the literature when constraining scattering amplitudes from data, although this can sometimes be justified in isolated cases.

\subsection{Partial-wave projection}

Typically the way analysis of scattering data proceeds is through a partial-wave analysis which starts by expanding the scattering amplitude in partial waves and analyzing the partial-wave projected amplitudes independently.
We focus on the case of two-to-two scattering, as this is much simpler and scattering more than two particles in an experiment is practically very challenging.
Scattering of two particles can still, of course, result in final states with more than two particles, but we do not address this here.
We will see in later sections that this is one advantage of lattice QCD, in that creating initial states with more than two particles is not a serious barrier.
In the case of two-to-two scattering of particles with helicity denoted by $\lambda$, the partial-wave expansion takes the form
\begin{equation}
    \mathcal{T}^{a_f a_i}_{\lambda^f_1 \lambda^f_2 ; \lambda^i_1 \lambda^i_2} (s, t) = \frac{\sqrt{s}}{\pi \sqrt{k_{a_i} k_{a_f}}} e^{i(\lambda_i - \lambda_f)\phi} \sum_{J} (2J+1) \; d^{(J)}_{\lambda^i \lambda^f} (\theta) \; t^{(J) a_f a_i}_{\lambda^f_1 \lambda^f_2 ; \lambda^i_1 \lambda^i_2} (s) ,
\end{equation}
where $\lambda^i = \lambda^i_1 - \lambda^i_2$, $\lambda^f = \lambda^f_1 - \lambda^f_2$, $a_i$ and $a_f$ denote the initial and final scattering channels, $k_{a_i}$ and $k_{a_f}$ are the magnitudes of the three-momentum of each particle in the initial and final states, $d_{\lambda^i \lambda^f}^{(J)}(\theta)$ is the Wigner (small) d-matrix, $t^{(J) a_f a_i}_{\lambda^f_1 \lambda^f_2 ; \lambda^i_1 \lambda^i_2} (s)$ are the helicity partial-wave amplitudes, and recall the relationship between $t$ and $\theta$ given in \Cref{eq:mandelstam}.
Note, however, that when not specifying a particular basis, the term partial-wave amplitude typically refers to the amplitudes in the $LS$-coupled basis, rather than the helicity basis used here.
The reason for this is that helicity states do not in general have definite parity, whereas states in the $LS$ basis do, and as states in the hadronic spectrum are specified with both total angular momentum $J$ and parity $P$ (i.e. $J^P$), it is convenient to default to a basis in which the states have definite parity.
However, the theoretical advantage of using the helicity basis is that for two-particle states the total spin component along the direction of relative motion is simply the difference of the helicities of the two particles, and therefore this difference is also the total angular momentum component along the direction of relative motion since the orbital angular momentum is always perpendicular to this direction.
Of course, simple relations between the helicity basis and the $LS$ basis can be used to easily convert between them.
In the case of spinless particles only, the sum over $J$ is replaced by a sum over the orbital angular momentum $\ell$ and the Wigner $d^J$ matrix is replaced by the $\ell$th Legendre polynomial.
In this simplified case, the helicity and $LS$ basis are trivially related.

When dealing with situations in which multiple two-particle scattering channels are kinematically allowed, it can be convenient to introduce the $K$-matrix defined by
\begin{equation}
    K_{a_f a_i}^{-1} (s) = 2T_{a_f a_i}^{-1} (s) + I_{a_f a_i} (s),
\end{equation}
where $I_{a_f a_i}(s) \equiv i \delta_{a_f a_i} \theta(s - s^{a_i}_{\rm thr})$, and $s^{a_i}_{\rm thr}$ is the location of $a_i$th scattering-channel threshold.
It can be shown that Hermiticity of the $K$-matrix is a general way to enforce unitarity of the $S$-matrix.
The convenience of this is simply the ease of constructing a Hermitian matrix over a unitary matrix.
Thus, one can simply parameterize the $K$-matrix such that it is Hermitian, and this will guarantee that unitarity of the $S$-matrix is satisfied.
Further, when below the three-particle production threshold, the $S$-matrix projected to definite angular momentum $J$, denoted here by $S^{(J)}$ and defined in terms of the partial-wave amplitudes $t^{(J)}$ by $S^{(J)} \equiv 1 + t^{(J)}$, is also itself unitary.
We can then define the $K$-matrix projected to definite angular momentum $J$ as $K^{(J)-1} \equiv 2 t^{(J)-1} + I$.
It is this partial-wave projected $K$-matrix, $K^{(J)}$, that is typically most convenient to work with.
It should also be noted that all that is really required to satisfy unitarity is for ${\rm Im}\, I_{a_f a_i}(s) = \delta_{a_f a_i} \theta(s - s^{a_i}_{\rm thr})$.
Therefore, one can use different functions for the real part of $I(s)$, some of which can improve the analytic behavior of the $K$-matrix.
For instance, the use of the Chew-Mandelstam function~\cite{Chew:1960iv} for $I(s)$ has been shown to remove unphysical singularities that can arise below threshold.
This procedure has been used in lattice QCD calculations, e.g. in Ref.~\cite{Wilson:2014cna} where further details about this prescription can be found.

\subsection{Experimental Analysis}

First, let us consider the case of elastic scattering of two spinless particles in which $S^{(J)}$ becomes a scalar and $J = \ell$.
We can then introduce the scattering phase shift $\delta_\ell$, defined via $S^{(\ell)} = e^{2i\delta_\ell}$ which leads to $K^{(\ell)-1} = \cot \delta_\ell$.
A common parameterization used to described the "bump" behavior in the cross sections coming from resonances is the Breit-Wigner (BW), defined by
\begin{equation}
    \tan \delta_\ell (s) = \frac{\Gamma_{\rm BW} /2}{m_{\rm BW} - \sqrt{s}} ,
\end{equation}
where $\Gamma_{\rm BW}$ is the BW width and $m_{\rm BW}$ is the BW mass of the resonance.
We can also now consider multiple spinless scattering channels, which adds a channel index to $S^{(\ell)}$.
For example, in the case of two scattering channels, we have
\begin{equation}
    S_{11}^{(\ell)} = \eta_\ell e^{2i \delta_\ell^{(1)}} , \;\;\; S_{22}^{(\ell)} = \eta_\ell e^{2i \delta_\ell^{(2)}} , \;\;\; S_{12}^{(\ell)} = S_{21}^{(\ell)} = i\sqrt{1 - \eta_\ell^2} e^{i(\delta_\ell^{(1)} + \delta_\ell^{(2)})} ,
\end{equation}
where $0 \leq \eta_\ell \leq 1$ is the inelasticity.
Another common parameterization, based on the multi-channel generalization of the effective range expansion, is
\begin{equation}
    k_{a_f}^{\ell+1/2} \; K^{(\ell)-1}_{a_f a_i} (s) \; k_{a_i}^{\ell+1/2} = \sum_{n=0} c^{n;\ell}_{a_f, a_i} \, s^n ,
\end{equation}
where the coefficients $c^{n;\ell}_{a_i, a_f}$ can be determined from fits to data, and some of these coefficients have names (e.g. $c^{0;\ell}$ is inversely related to the scattering length).
Finally, in some cases it is possible to generalize to scattering involving particles with spin, but this typically requires the particular system to allow for a basis in which $S^{(J)}$ is diagonal in the spin indices, which can be done in e.g. $N \pi$ and nucleon-nucleon scattering.
Note that in this case, the $K$-matrix projected to definite angular momentum $J$, $K^{(J)}$, would then include indices to describe the extra degrees of freedom coming from spin.
The parameterizations mentioned here are not exhaustive by any means, and many others have been considered in the literature.

However, in general, any given parameterization may rely on certain assumptions.
For instance, the BW parameterization assumes the resonance in question is both narrow and isolated from other states,
but exactly how narrow and isolated the resonance must be is not easily defined.
Further, using the BW parameters to characterize a resonance are not in general independent of the particular process being studied.
The misuse of the BW parameterization in this way has led to much confusion regarding states in which these assumptions are far from valid, like in the case of the $\sigma$ resonance.
One really should instead aim to use parameterizations which satisfy the analytic constraints, and then analytically continue the resulting scattering amplitude to search for poles corresponding to the resonance. 

\begin{figure}
    \centering
    \includegraphics[width=.5\columnwidth]{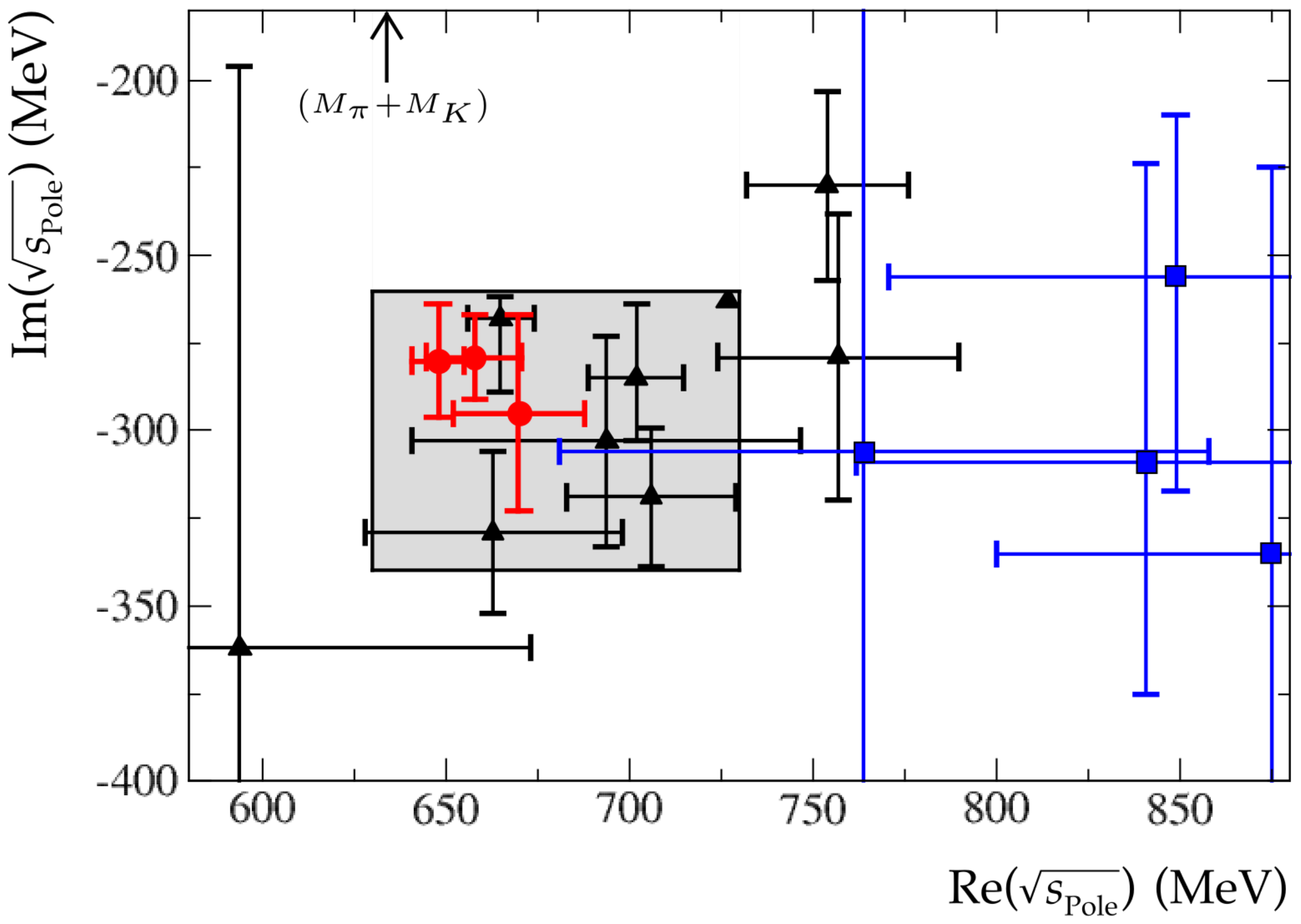}
    \caption{Spread of pole positions for the $\sigma$ resonance from various analyses.
    The red circles are from analyses that use dispersive approaches, while blue squares are from BW fits.
    Figure taken from Ref.~\cite{Workman:2022ynf}.}
    \label{fig:sigma_pole}
\end{figure}

The analytic structure demanded for the scattering amplitude can be implemented in a consistent manner that respects crossing symmetry through a set of integral equations, referred to as dispersion relations.
These techniques, first used for $\pi\pi$ scattering, were pioneered by Roy~\cite{Roy:1971tc}.
However, several analyses did not use these dispersive techniques, due to the simplicity of applying strategies like the BW parameterization.
And, before these techniques became widely used, there was a long history of confusing and inconsistent results surrounding the properties of the $\sigma$ resonance.
Given the relevance of the $\sigma$ in nucleon interactions, this was an important puzzle to resolve.
The problem, as has been alluded to, was coming from the location of the $\sigma$ resonance pole lying far enough into the complex plane that the use of the BW to describe the $\pi\pi$ scattering data was not leading to a clean picture.
As is shown in \Cref{fig:sigma_pole}, the pole positions coming from dispersive analyses lead to much more stable pole positions.
For a detailed overview of this long history, see the review in Ref.~\cite{Pelaez:2015qba}.

Given the difficulties in extracting pole positions deep in the complex plane, especially if several such poles are all nearby, and that typical analyses of scattering data do not take into full consideration the correct analytic structure of the scattering amplitudes, this leads one to ask how certain we are of the current state of the resonances found in experiment.
Of particular interest is how reliably the huge number of exotic candidates, which call into question the validity of the quark model, have been determined.

For example, the COMPASS experiment had confirmed earlier claims of two hybrid mesons determined from $\eta \pi$ and $\eta^\prime \pi$ final states, namely the $\pi_1(1400)$~\cite{COMPASS:2014vkj} and the $\pi_1(1600)$~\cite{COMPASS:2018uzl}.
However, a recent reanalysis of the data by the JPAC collaboration using more robust methods found evidence for only one $\pi_1$ hybrid meson~\cite{JPAC:2018zyd}, which has since been confirmed from a lattice calculation~\cite{Woss:2020ayi}.

There are also other situations that can give the appearance of a resonance, like triangle singularities~\cite{COMPASS:2020yhb} and threshold cusps~\cite{Swanson:2015bsa}.
Therefore it is of great importance to the field that we strive to use modern analysis methods that take into account the relevant theoretical constraints on the scattering amplitudes in order to confirm the resonances being claimed in various experiments.
This includes using both experimental and lattice methods.

\section{Scattering on the Lattice}

As we have seen, much of the knowledge of the spectrum of hadrons comes from scattering experiments involving the preparation of asymptotic states which can occupy a continuum of states by varying the initial momenta followed by real-time dynamics as they become close enough to interact and finally the measurement of the resulting asymptotic states.
Conversely, lattice QCD simulations are performed in a finite, Euclidean volume where the eigenstates become discrete and asymptotic states cannot be created.
At first glance, the differences between the two approaches appear significant enough to question whether lattice QCD is able to provide scattering amplitudes in a comparable fashion to experiment.
Indeed, there are two main obstacles to consider: the use of a Euclidean metric (i.e. an imaginary-time setup) and the restriction to a finite volume where asymptotic states cannot be constructed.
The difficulty of the first issue was investigated by Maiani and Testa~\cite{Maiani:1990ca} where it was observed that Euclidean correlators of temporally-separated pion fields could only give access to the scattering amplitude at one energy, namely at threshold.
While, in principle, analytic continuation to Minkowski space could resolve this issue, using lattice data to do this leads to an ill-posed inverse problem.
However, there has been recent work to revisit these approaches through the extraction of the relevant spectral functions~\cite{Hansen:2017mnd,Bulava:2019kbi}, including extensions of the results from Maiani and Testa to go beyond threshold by utilizing modified correlators~\cite{Bruno:2020kyl}.
These methods, while addressing some of the limitations of current methods, typically need larger volumes than are used in most lattice simulations in order to neglect finite-volume corrections.
This brings us to the second obstacle for lattice simulations, namely the finite volume.

The approach introduced by L\"uscher~\cite{Luscher:1985dn,Luscher:1986pf} in fact uses the finite volume as a tool rather than a hindrance.
Similarly to the spectral function approaches mentioned above, this method relies on the extraction of a metric-independent observable.
In L\"uscher's approach, this observable is the two-particle finite-volume spectrum which can be shown to have its finite-volume dependence dictated by the interactions of the two particles and thereby providing constraints on the corresponding scattering amplitude, referred to as the quantization condition.
The intuition for this is that the finite volume prevents asymptotic states or, put another way, it forces the particles in the box to interact thereby shifting the energy of the multi-particle state away from that of just multiple free particles.
Although not the focus here, it is important to mention another alternative to L\"uscher's approach, known as the HAL QCD potential method~\cite{Ishii:2006ec,Ishii:2012ssm} which has been used extensively by the HAL QCD collaboration.
The basic idea is to extract the Nambu-Bethe-Salpeter (NBS) wave function from correlators involving two spatially-separated single-particle operators along with a two-particle operator to create the two-particle states of interest.
Assuming the two-particle states created only contain insignificant contributions from inelastic states, the NBS wave function can be used to extract a potential and then scattering observables from using this potential to solve the Schr\"odinger equation.
It should be noted that the potentials extracted in the HAL QCD method rely on a derivative expansion which must be truncated in practical calculations.
A review on the HAL QCD method, including a discussion on the systematics associated with this truncation can be found in Ref.~\cite{Aoki:2020bew}.
This is not to say the L\"uscher method does not have its own systematic errors to deal with, and we will discuss these in turn in the following sections.

\subsection{Finite-Volume spectrum}

An essential aspect in applications of the L\"uscher formalism and its extensions is the extraction of the relevant finite-volume spectrum.
This is important because in many studies it is where the largest systematic errors can enter the analysis.
Therefore, we discuss it in detail here.
In principle, one can extract the finite-volume spectrum from a two-point temporal correlation function of the form
\begin{equation}
\begin{split}
    C(t_{\rm sep}) &= \sum_{t_0} \braket{\mathcal{O}_{\rm snk} (t_{\rm sep} + t_0) \mathcal{O}^\dagger_{\rm src}(t_0)} \\
    &= \sum_{n=0}^\infty \bra{\Omega} \mathcal{O}_{\rm snk} \ket{n} \bra{\Omega} \mathcal{O}_{\rm src} \ket{n}^\ast e^{-E_n t_{\rm sep}} ,
\end{split}
\end{equation}
where $\ket{\Omega}$ denotes the vacuum state, $\mathcal{O}_{\rm snk}(t)$ and $\mathcal{O}_{\rm src}(t)$ are generic interpolating operators, $t_{\rm sep}$ is the time separation between the operators, $\ket{n}$ denotes the $n$th eigenstate of the Hamiltonian corresponding to the energy $E_n$, and thermal effects have been ignored.
The difficulty in extracting the spectrum from a correlator like this is merely a practical one coming from the finite statistics and the exponentially bad signal-to-noise ratio, which e.g. for baryons of mass $m_B$ scales as $e^{-(m_B - 3 m_\pi/2)t_{\rm sep}}$.
The finite statistics can make fits including more than two or three exponentials difficult, and if one instead truncates the fit model to the first few eigenstates then this necessitates $t_{\rm sep}$ to be large enough to make the model valid which in turn leads to an exponentially worse signal.
Therefore, this can lead to a large systematic error in the extracted spectrum due to contamination from higher excited states, including elastic states that do not cause issues in the HAL QCD method.

To avoid these issues, the common approach is to construct a Hermitian correlator matrix of the form
\begin{equation}
    C_{ij} (t_{\rm sep}) = \sum_{t_0} \braket{\mathcal{O}_i (t_{\rm sep} + t_0) \mathcal{O}_j^\dagger (t_0)} ,
\end{equation}
where the set of interpolating operators $\left\{\mathcal{O}_i\right\}$ is chosen such that each operator has overlap with at least one of the desired states.
It was demonstrated that the eigenvalues of this matrix behave as
\begin{equation}
    \lambda_n (t_{\rm sep}) \propto e^{-E_n t_{\rm sep}} + O(e^{-\Delta_n t_{\rm sep}}) ,
\end{equation}
where $\Delta_n \equiv \rm{min}_{m \neq n} |E_n - E_m|$~\cite{Luscher:1990ck}, which offers a method for extracting several excited states, but the contamination in these eigenvalues coming from other states can be significant for dense spectra.
It was proposed that forming a generalized eigenvalue problem (GEVP)
\begin{equation}
    C(t_{\rm sep}) \upsilon_n (t_{\rm sep}, \tau_0) = \lambda_n(t_{\rm sep}, \tau_0) C(\tau_0) \upsilon_n (t_{\rm sep}, \tau_0) 
\end{equation}
could ameliorate the situation.
This was investigated in detail where it was found that for $\tau_0$ chosen large enough, the leading contamination to the generalized eigenvalues could be lessened with $\Delta_n = E_N - E_n$ where $N$ is the size of the correlator matrix~\cite{Blossier:2009kd}.
Another advantage of this method is the coefficients appearing in front of the exponentials in the eigenvalues are positive, meaning the effective energy of $\lambda_n(t_{\rm sep}, \tau_0)$ monotonically approaches $E_n$.
This means that determining ground-state saturation is more reliable when a plateau in the effective energy is observed, whereas a plateau from correlators with both positive and negative coefficients could be the result of delicate cancellation coming from contamination in the eigenvalues.

As a closing remark, the importance of the operator set should be emphasized.
Situations can arise in which the extracted spectrum has non-negligible dependence on the set of operators used.
For example, if the chosen operator set has very small, but non-zero, overlap onto some states in the region of interest, this can lead to an incorrect extraction of the spectrum.
This dependence has been observed in lattice calculations, e.g. in the study of the $\rho$ resonance~\cite{Wilson:2015dqa}.

This question is of particular importance to the long-standing two-nucleon disagreement in the literature, which is exacerbated by the poor signal-to-noise ratio for baryons.
Several modern studies of two-baryon systems using the L\"uscher method have used the GEVP to lessen these issues~\cite{Francis:2018qch,Horz:2020zvv,Green:2021qol,Amarasinghe:2021lqa}.
These works have led to new understandings of the problems,
but further studies are needed before the disagreement can be fully resolved.
For some perspective on this as it applies to two-baryon systems, see Ch. 15 and 16 of Ref.~\cite{Tews:2022yfb}.

\subsection{Two-particle Quantization Condition}

In the approach first introduced by L\"uscher, the derivation was for two identical spinless particles with zero total momentum.
Since that time, this has been generalized to include non-zero momentum~\cite{Rummukainen:1995vs,Kim:2005gf,Christ:2005gi}, multiple scattering channels~\cite{Lage:2009zv,Bernard:2010fp,Doring:2011vk}, and non-degenerate particles~\cite{Fu:2011xz,Hansen:2012tf} with intrinsic spin~\cite{Gockeler:2012yj,Briceno:2012yi,Briceno:2014oea}.
For a recent review, see Ref.~\cite{Briceno:2017max}.
In summary, the quantization condition, valid up to the first threshold including three or more particles and ignoring exponentially suppressed corrections in the volume, takes the form
\begin{equation}
\label{eq:2QC}
    \det [\tilde{K}^{-1} (E_{\rm cm}) - B^{\bm{P}} (E_{\rm cm})] = 0 ,
\end{equation}
where $B^{\bm{P}} (E_{\rm cm})$ is a known finite-volume kinematic function called the box matrix,
and the angular momentum projections of $\tilde{K}$ are defined by 
\begin{equation}
    \tilde{K}^{(J) -1}_{\ell_f S_f a_f; \ell_i S_i a_i} (E_{\rm cm}) = k^{\ell_f + 1/2}_{a_f} K^{(J) -1}_{\ell_f S_f a_f; \ell_i S_i a_i} (E_{\rm cm}) k^{\ell_i + 1/2}_{a_i} ,
\end{equation}
where the $\ell$ and $S$ indices correspond to the orbital angular momentum and spin, respectively.
The notation used here follows that in Ref.~\cite{Morningstar:2017spu},
but many other equivalent forms have been utilized, e.g. in Ref.~\cite{Woss:2020cmp}.
When used in practical calculations, where the energies extracted correspond to irreducible representations (irreps) of the reduced symmetry group for a lattice, the determinant condition is block diagonalized in these irreps and one can focus on the corresponding block for any given energy.
Further, owing to the partial-wave mixing induced by the reduced symmetry of the lattice,
the blocks are still of infinite dimension and one must truncate the quantization condition at a maximum value $\ell_{\rm max}$.
The truncation errors are usually observed to be small, due to the angular-momentum barrier, however this can also lead to difficulty in constraining the higher partial waves that contribute.

The general procedure for the practical use of the quantization condition is to parameterize $\tilde{K}$ (or $\tilde{K}^{-1}$) and adjust the free parameters until the spectrum predicted from the quantization condition for these parameters (determined by finding all values of $E_{\rm cm}$ that lead to a vanishing determinant) matches the spectrum extracted from the lattice calculation.
The use of this quantization condition is at a very mature level, having been used extensively by the community.

\subsubsection{Left-hand cuts}

There is another limitation of the two-particle quantization condition coming from allowed particle exchanges in the $u$ and $t$ channels.
Specifically, these particle exchanges lead to poles in the scattering amplitude that become cuts, commonly referred to as left-hand cuts, after projection to a particular partial wave.
These cuts induce an imaginary part of the scattering amplitude which the current form of the quantization condition is not capable of reproducing.
During much of the work on developing the quantization condition, it was not expected that this limitation would pose a problem, as many systems do not produce energies near the left-hand cuts.
However, in cases where the scattering particles are much heavier than the exchanged particles in the crossed channels, finite-volume energies can become quite near and even cross the start of these cuts.
This has been seen already in some lattice QCD studies~\cite{Green:2021qol,Padmanath:2022cvl}.

Recently, there have been various proposals for circumventing the issue by avoiding the partial wave projections~\cite{Meng:2023bmz}, and even extensions of the quantization condition below the start of the first left-hand cut~\cite{Raposo:2023oru}.
For systems in which one of the scattering particles is a two-body bound state, there is also a third option, which is to utilize the three-particle quantization condition which has recently been demonstrated to appropriately handle the issues arising in the two-particle quantization condition~\cite{Dawid:2023jrj}.
These studies have also revealed issues when using energies that are simply too close to, but still above, the left-hand cut where the exponentially-suppressed volume corrections ignored in the derivation can become large.

\subsection{Three-particle Quantization Condition}

The majority of resonances have open decay channels involving three or more particles, and therefore the limits of the two-particle quantization condition prevent their study.
Early work towards the extension of the quantization condition beyond the first three-particle threshold was done in Refs.~\cite{Polejaeva:2012ut,Briceno:2012rv}.
Soon after, the first fully-relativistic three-particle quantization condition making no approximations regarding the two-particle interactions, referred to as the relativistic field theory (RFT) approach, was worked out~\cite{Hansen:2014eka}, which takes the form
\begin{equation}
    \det [\mathcal{K}_{\rm df,3} (E^\ast) + F_3(E, \bm{P}, L)^{-1}] = 0 ,
\end{equation}
where $\mathcal{K}_{\rm df,3}$ is the divergence-free three-particle $K$-matrix used to describe the short-distance three-particle interactions, and $F_3$ contains both kinematical finite-volume functions as well as the two-particle $K$-matrix.
Although the form of this quantization condition is reminiscent of the two-particle quantization condition in \Cref{eq:2QC}, there are some key differences.
First there is no clean separation between finite and infinite volume physics anymore since $F_3$ depends on the interactions in the two-particle subchannels through its dependence on the two-particle $K$-matrix, and second $\mathcal{K}_{\rm df,3}$ is an intermediate scheme-dependent quantity.
Obtaining the three-to-three scattering amplitude requires the solution of integral equations~\cite{Hansen:2015zga,Jackura:2020bsk,Dawid:2023jrj}.
One of the reasons for this introduction of the intermediate quantity $\mathcal{K}_{\rm df,3}$ is the need to truncate angular momentum decompositions of the scattering amplitudes in practical implementations of the formalism.
Specifically, long distance contributions to the three-to-three scattering amplitude coming from subchannel two-to-two interactions prevents the general uniform convergence of the angular momentum decomposition of the three-to-three scattering amplitude.
Thus, these long-distance contributions are formally subtracted off to produce the quantity $\mathcal{K}_{\rm df,3}$.

It should not be too surprising to see the two-particle $K$-matrix in the quantization condition, however, as the two-particle pairs within the three-particle system produce constraints on the two-particle interactions.
Therefore, typically one should perform a global fit to the two- and three-particle spectrum utilizing both the two- and three-particle quantization conditions in order to benefit from the constraints on the two-particle interactions within the three-particle system.
Typically, the parameterizations used for the three-particle interactions are determined by truncating the threshold expansion of $\mathcal{K}_{\rm df, 3}$ to some order.
Explicit details of this parameterization, and other details of implementing the three-particle quantization condition can be found in e.g. Ref.~\cite{Blanton:2019igq}.

Since the introduction of the RFT formalism, first derived for a system of three scalar fields without transitions to two particles, there have been several extensions.
These include allowing for $\bm{2} \leftrightarrow \bm{3}$ interactions~\cite{Briceno:2017tce}, allowing for two-particle subchannel resonances~\cite{Briceno:2018aml}, non-maximal isospin~\cite{Hansen:2020zhy}, non-degenerate particles~\cite{Blanton:2020gmf,Blanton:2021mih}, non-zero spin~\cite{Draper:2023xvu}, etc.
Further, returning to the issue of left-hand cuts in the two-particle quantization condition, the formalism for circumventing this in studies of the $T_{cc}^+$ with the three-particle quantization condition have been worked out~\cite{Hansen:2024ffk}.

There are also two other competing formalisms to the RFT approach, namely the non-relativistic effective field theory (NREFT) approach~\cite{Hammer:2017uqm,Hammer:2017kms} and the finite-volume unitarity (FVU) approach~\cite{Mai:2017bge}.
All three approaches are reviewed in Ref.~\cite{Hansen:2019nir}.
It has also been shown recently that the RFT and FVU approaches are equivalent~\cite{Jackura:2019bmu,Blanton:2020jnm,Jackura:2022gib}

With the development of the three-particle quantization conditions beginning to reach a mature level, there have been several lattice QCD studies of three-meson systems utilizing these formalisms, e.g. studies involving pions and kaons with maximal isospin~\cite{Blanton:2019vdk,Fischer:2020jzp,Brett:2021wyd,Blanton:2021llb}, three-body resonances~\cite{Mai:2021nul,Sadasivan:2021emk}, extractions of the three-to-three scattering amplitude~\cite{Hansen:2020otl}, etc.

\section{Recent Applications}

The use of the two-particle and three-particle quantization conditions and the HAL QCD method has reached a mature level, with hundreds of works performing these types of calculations.
Like in experiment, the two-particle quantization condition constrains the scattering amplitude on the real-axis only and one must infer the poles from this through analytic continuation.
Here we review some recent work in applying these methods, with an emphasis on recent studies.

\subsection{The two-pole structure of the \texorpdfstring{$\Lambda(1405)$}{Lambda(1405)}}

There has been much debate regarding the nature of the $\Lambda(1405)$ (see recent reviews in e.g. Refs.~\cite{Mai:2018rjx,Meissner:2020khl,Mai:2020ltx,Hyodo:2020czb}).
First, the fact that this resonance is lighter than its nucleon counterpart, despite having a strange quark, makes fitting this state into the quark model difficult~\cite{Isgur:1978xj}.
Second, the use of unitarized chiral perturbation theory (UChPT) produces strong support for the $\Lambda(1405)$ actually corresponding to two poles, first suggested in Ref.~\cite{Oller:2000fj}.
This second pole, referred to as the $\Lambda(1380)$ has recently made the listings by the Particle Data Group~\cite{Workman:2022ynf}, albeit with less confidence.

\begin{figure}
    \centering
    \includegraphics[width=.54\columnwidth]{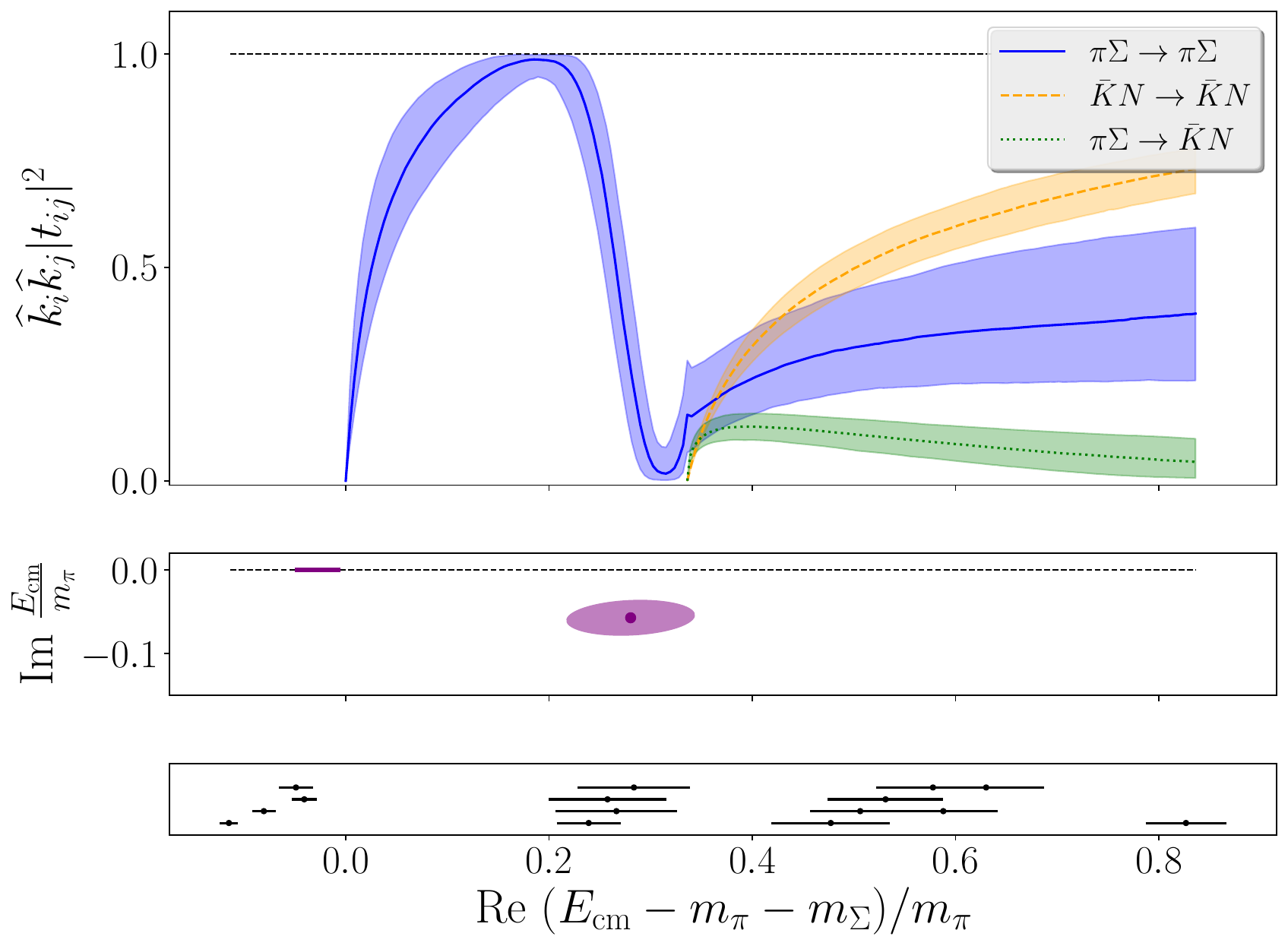}
    \caption{The squared scattering amplitudes (top), the pole locations (middle), and the finite-volume energies (bottom) for the $\pi\Sigma-\bar{K}N$ system from a lattice QCD calculation with $m_\pi \approx 200$ MeV~\cite{BaryonScatteringBaSc:2023ori}.}
    \label{fig:lambda_poles}
\end{figure}

There have been a few previous studies from lattice QCD of this resonance, but only recently has there been a calculation which included the relevant coupled channels~\cite{BaryonScatteringBaSc:2023zvt,BaryonScatteringBaSc:2023ori}.
The main result of this work is shown in \Cref{fig:lambda_poles}.
From the analytic continuation of the constrained scattering amplitude, two poles were found, one corresponding to a virtual bound state and the other a resonance.
Given that this calculation was performed at a pion mass of $m_\pi \approx 200$ MeV, this is in fact consistent with the predictions from UChPT.

\subsection{Poles of the \texorpdfstring{$D^\ast_0(2300)$}{D*0(2300)}}

Studies using UChPT have also suggested other resonances may in fact have a two-pole structure, with each pole corresponding to different SU(3) flavor representations~\cite{Meissner:2020khl}.
For example, in the charmed meson sector, there there has been much discussion raised regarding the nature of the $D^\ast_0(2300)$ (originally named the $D^\ast_0(2400)$) discovered by Belle in 2003~\cite{Belle:2003nsh}.
This resonance has a strange partner, the $D^\ast_{s0}(2317)$ discovered by BaBar, also in 2003~\cite{BaBar:2003oey}.
These were among the many states being seen in the early 2000s that did not fit predictions from the quark model.
In particular, why do the two states have very similar masses, given the expectation of a heavier $D^\ast_{s0}(2317)$ due to its strange valence quark?

\begin{figure}
    \centering
    \includegraphics[width=.54\columnwidth]{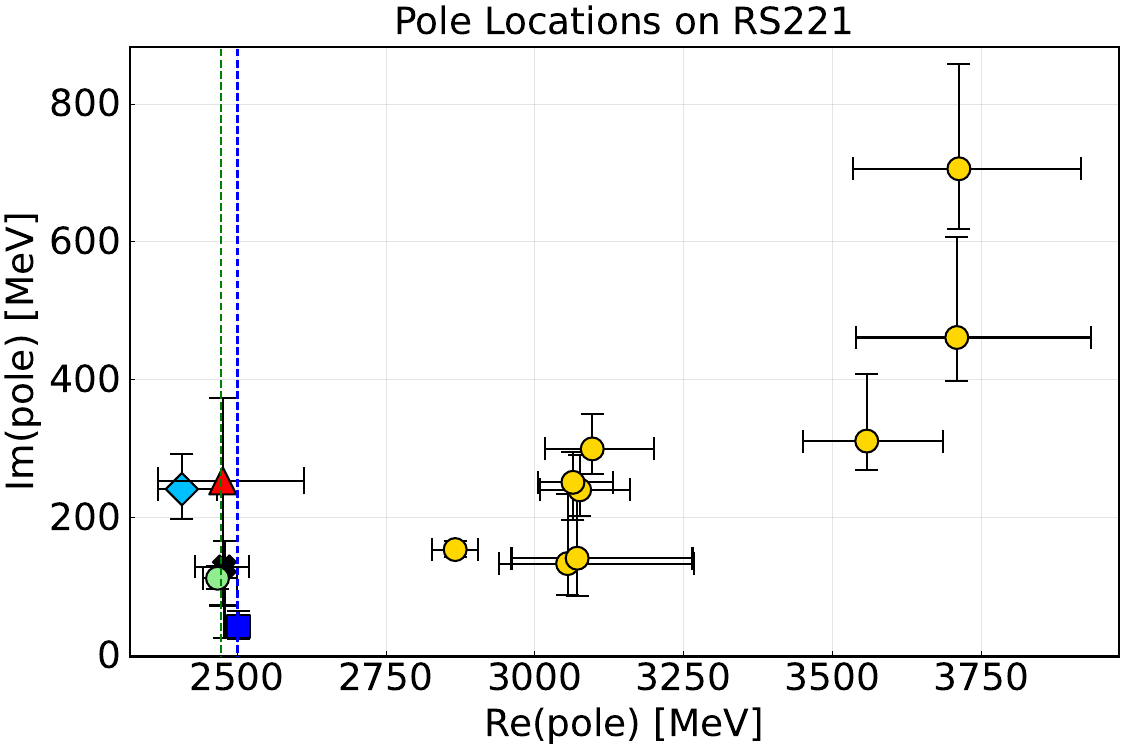}
    \caption{Second pole positions from the scattering amplitudes for the coupled $D\pi-D\eta-D_s \bar{K}$ system.
    The pole predicted from UChPT with scattering lengths obtained in lattice QCD is shown in green~\cite{Albaladejo:2016lbb}.
    The other poles come from the lattice data in Ref.~\cite{Moir:2016srx} without constraints from UChPT (yellow) and with constraints from UChPT~\cite{Asokan:2022usm}.
    The vertical dashed lines correspond to the $D\eta$ (green) and $D_s \bar{K}$ (blue) thresholds.}
    \label{fig:D0_poles}
\end{figure}

The extraction of these resonances was initially based on the Breit-Wigner model, which many argue is not a good approximation for this system.
Indeed, there is a consistent picture formed from the use of UChPT with input from lattice QCD~\cite{Liu:2012zya} which predicts a two-pole nature for the $D^\ast_0(2300)$~\cite{Albaladejo:2016lbb}.
However, the question arose as to why a second stable pole position was not found in the lattice QCD study in this system from the Hadron Spectrum Collaboration~\cite{Moir:2016srx}.
Of course, this is not all that unexpected, as the second pole is predicted to be on a Riemann sheet that is not directly connected to the physical sheet where stable pole predictions can be challenging.
Additionally, poles near threshold with a small residue can produce effects similar to that of a pole far from threshold with a larger residue, further exacerbating the problem of finding stable pole positions.
This question was recently revisited with a reanalysis of the lattice data, but including constraints from UChPT which led to much more stable second pole positions that are consistent with predictions from UChPT~\cite{Asokan:2022usm}.
The locations of these poles are shown in \Cref{fig:D0_poles}.

\subsection{The \texorpdfstring{$\sigma$}{sigma} Resonance via Dispersion Relations}

Most lattice calculations to date only explicitly enforce the unitarity of the $S$-matrix, but ignore the left-hand cuts associated with crossing symmetry.
This is generally accepted, as inclusion of the full analytic structure of the scattering amplitude is not strictly necessary in certain circumstances.
However, for poles lying near a left-hand cut or poles lying deep in the complex plane such that left-hand cuts can produce competing effects for the scattering amplitude on the physical sheet, it is typically necessary to include these cuts in the description of the scattering amplitude.

\begin{figure}
    \centering
    \includegraphics[width=.54\columnwidth]{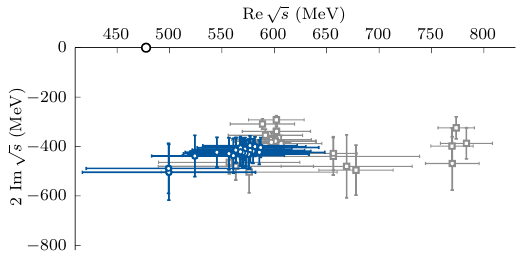}
    \caption{The set of pole locations for the $\sigma$ resonance extracted from different parameterizations for the scattering amplitude.
    The blue points correspond to parameterizations that obey the dispersion relations, while those in gray do not.
    The calculation was performed at a pion mass $m_\pi \approx 239$ MeV~\cite{Rodas:2023twk}.}
    \label{fig:sigma_pole_lattice}
\end{figure}

As discussed above, one way to consistently implement the correct analytic structure is through the use of dispersion relations.
This was recently done in a lattice QCD calculation of all three isospin $\pi \pi$ channels~\cite{Rodas:2023gma,Rodas:2023twk}.
The dispersion relations in that work were implemented with
\begin{equation}
    \tilde{t}^{(\ell)}_I (s) = \tau^{(\ell)}_I (s) + \sum_{I^\prime, \ell^\prime} \int_{4 m^2_\pi}^{\infty} d s^\prime K_{\ell \ell^\prime}^{II^\prime} (s^\prime, s) \; {\rm Im} \; t^{(\ell^\prime)}_{I^\prime} (s^\prime) ,
\end{equation}
where $t^{(\ell)}_I$ is the partial-wave projected scattering amplitude with isospin $I$, $\tau^{(\ell)}_I$ are polynomials in $s$ with parameters determined by the number of subtractions used,\footnote{These subtractions included in the dispersion relations are used to ensure the contour at infinity can safely be ignored.}
and $K_{\ell \ell^\prime}^{II^\prime} (s^\prime, s)$ are known functions depending on the number of subtractions.
To make use of these dispersion relations, first a set of scattering amplitudes that only include the constraints from unitarity are considered and fed into the right-hand-side of the equation.
Then, those that satisfy $\tilde{t}^{(\ell)}_I (s) = t^{(\ell)}_I (s)$ can be shown to implement crossing symmetry and analyticity correctly.
As was found when analyzing experimental data, the amplitudes obeying these properties lead to much more stable pole positions, as seen in \Cref{fig:sigma_pole_lattice}.

\subsection{The doubly-charmed tetraquark \texorpdfstring{$T^+_{cc}(3875)$}{Tcc+(3875)}}

The recently observed $T_{cc}^+$ by LHCb~\cite{LHCb:2021vvq,LHCb:2021auc} in the $D^0D^0\pi^+$ mass spectrum just below the $D^{\ast +}D^0$ threshold has received a lot of attention, as it must be an exotic hadron given the two valence charm quarks and overall integer spin.
Further interest has been generated in its uniquely narrow width, as compared to most other exotic hadron candidates observed.
There were a few investigations of this system using lattice QCD prior to the discovery, including a study utilizing the HAL QCD method which found no evidence of a bound state or resonance for the pion masses considered~\cite{Ikeda:2013vwa} and studies of the finite-volume spectrum~\cite{Cheung:2017tnt,Junnarkar:2018twb} which did not rule out such a state but did not extract a pole position.

\begin{figure}
    \centering
    \includegraphics[width=.9\columnwidth]{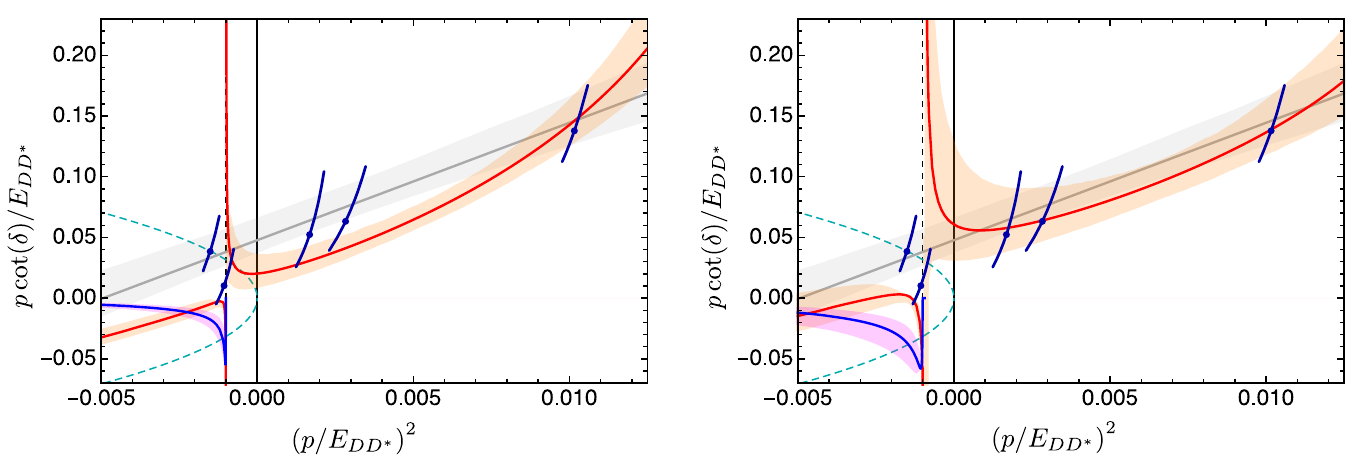}
    \caption{The scattering phase shift for the $DD^\ast$ system as a function of the scattering momentum $p$.
    The energies were determined in Ref.~\cite{Padmanath:2022cvl}, and the gray band corresponds to the fit performed in that work.
    The orange (real part) and purple (imaginary part) are from fits done in Ref.~\cite{Du:2023hlu} taking into account the effects from the left-hand cut (which starts from the vertical dashed line).
    The left figure uses all data (including errors) above the left-hand cut, while the right figure only uses the datapoints that lie entirely above the left-hand cut.}
    \label{fig:tcc_left-hand-cut}
\end{figure}

Naturally, the discovery by LHCb led to new studies of its existence from lattice QCD.
The first such study utilizing the two-particle quantization condition found a virtual bound state in the $DD^\ast$ scattering amplitude~\cite{Padmanath:2022cvl}.
Given the larger than physical pion mass $m_\pi \approx 280$ MeV, this may indeed correspond to the $T^+_{cc}$ which could evolve from a resonance to a virtual bound state as the pion mass is increased.
However, it was later pointed out that some of the energies used were near or below the first left-hand cut which leads to uncontrolled systematics for phase shifts coming from those energies~\cite{Du:2023hlu}.
In this work, a reanalysis was done with the lattice data, taking into account the constraints from the left-hand cuts.
The results are shown in \Cref{fig:tcc_left-hand-cut}.
Additionally, given the proximity of the left-hand cut, this should be taken into account in the scattering amplitudes utilized. Overall, this limits the conclusions that can be drawn.
There were also recent lattice QCD results for $m_\pi \approx 350$ MeV, that extracted a single energy level near threshold which, however, does not allow for a reliable constraint on the scattering amplitude~\cite{Chen:2022vpo}.
Finally, there was another study using the HAL QCD method with a nearly physical pion mass which also observed a virtual bound state~\cite{Lyu:2023xro}.
However, the effects of the left-hand cut were also not considered in this study.
The three-particle quantization condition has yet to be used to solve these issues, but, as the formalism for this is now available, results for this should soon be available.

\subsection{Three-meson scattering}

\begin{figure}
    \centering
    \includegraphics[width=.56\columnwidth]{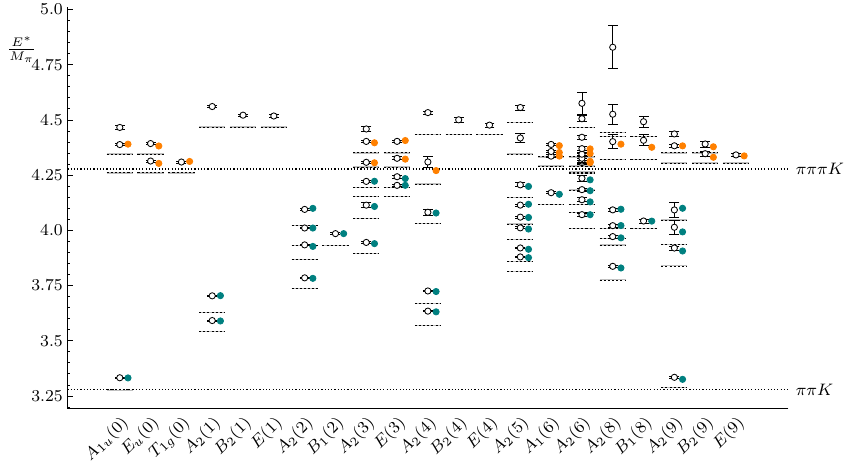}
    \includegraphics[width=.43\columnwidth]{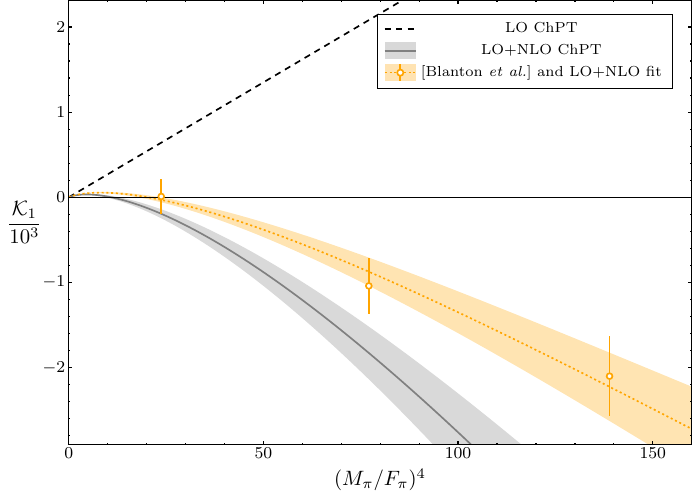}
    \caption{(Left) The $\pi\pi K$ finite-volume spectrum from lattice QCD with $m_\pi \approx 340$ MeV~\cite{Draper:2023boj}.
    The colored points correspond to predictions from the three-particle quantization condition with the parameterizations constrained by the energies associated with the teal points.
    The orange points were not included in the fits.
    (Right) The leading-order (LO) and next-to-leading order (NLO) predictions of $\mathcal{K}_1$ for a $\pi\pi\pi$ system~\cite{Baeza-Ballesteros:2023ljl} compared with results from lattice QCD~\cite{Blanton:2021llb}.}
    \label{fig:3QC}
\end{figure}

As discussed above, there have now been several numerical studies of three-meson systems using lattice QCD.
The past year saw the first application of the RFT approach to mixed systems of pions and kaons~\cite{Draper:2023boj}, utilizing the formalism and implementation details presented in Refs.~\cite{Blanton:2021mih} and~\cite{Blanton:2021eyf}, respectively.
The $\pi\pi K$ spectrum for one ensemble in this work is shown on the left in \Cref{fig:3QC}.
These results, along with those in Ref.~\cite{Blanton:2021llb}, include constraints on higher-partial waves and give strong evidence for significant short-range three-particle interactions in $\mathcal{K}_{\rm df,3}$ which have been ignored in many studies.
This work also made use of Wilson ChPT~\cite{Sharpe:1998xm,Bar:2003mh} to make an assessment of the leading-order systematics errors induced by the non-zero lattice spacing, which were found to be negligible within statistics.

Furthermore, this year we saw the next-to-leading order calculations of $\mathcal{K}_{\rm df,3}$ for three pions in ChPT~\cite{Baeza-Ballesteros:2023ljl}, which showed surprisingly strong corrections to the leading order results, relieving the concerns of the strong disagreement between the leading-order ChPT expectations for the lowest two terms in the threshold expansion initially seen in Ref.~\cite{Blanton:2021llb}.
These results for the coefficient $\mathcal{K}_1$ of the second term in the threshold expansion, along with the lattice data of Ref.~\cite{Blanton:2021llb} are shown on the right in \Cref{fig:3QC}.
This work has recently been extended to all isospin channels for three pions~\cite{Baeza-Ballesteros:2024mii}.

\section{Conclusions and Outlooks}

There has been significant progress towards more robust determinations of the hadronic spectrum from both lattice QCD and the wider community.
Here we saw the extraction of pole positions, the implementation of dispersive techniques, progress on dealing with left-hand cuts, exploration of three-hadron systems, etc.

However, there is still a significant amount of work to be done.
We are far from a complete understanding of the spectrum of QCD.
Some of the problems within hadron spectroscopy, and potential directions toward their solutions, were reviewed.
We focused on the questions that lattice QCD can address.

It is clear from the current work being done that calculations involving some of the puzzles within the hadronic spectrum, like robust calculations of the $T_{cc}^+$, the Roper resonance, etc., are on the horizon.
It is truly an exciting time to be working in hadron spectroscopy with lattice QCD.

\acknowledgments

I thank the organizers of the Lattice 2023 conference for giving me the opportunity to give a plenary.
I also thank Erin Blauvelt and Ben H\"orz for their feedback on these proceedings.
This work is supported by
the U.S. Department of Energy, Office of Science, Office of Nuclear Physics through \textit{Contract No. DE-SC0012704}, within the framework of Scientific Discovery through Advanced Computing (SciDAC) award \textit{Fundamental Nuclear Physics at the Exascale and Beyond}, and by the U.S. National Science Foundation (NSF-OAC-2311430).

\bibliographystyle{JHEP}
\bibliography{refs}

\end{document}